\documentstyle[12pt,graphicx]{article}
\title{Emergent physics on vacuum energy and cosmological constant
\footnote{submitted to Proceedings of the 24th International Conference on
Low Temperature Physics, the pdf file with the viewgraphs for
oral presentation is in
http://ltl.tkk.fi/personnel/THEORY/volovik/LT-volovik.pdf}}
\author{G.E. Volovik
\\Low Temperature Laboratory, Helsinki University of
Technology\\
P.O.Box 2200, FIN-02015 HUT, Finland\\
\\
 L.D. Landau Institute for
Theoretical Physics\\  Kosygin Str. 2, 119334 Moscow, Russia
}

\begin{document}
\maketitle

\begin{abstract}
{ The phenomenon of emergent physics in condensed-matter many-body
systems has become the paradigm of modern physics, and can probably also
be applied to high-energy physics and cosmology. This encouraging
fact comes from the universal properties of the ground state (the analog of
the quantum vacuum) in fermionic many-body systems, described in
terms of the momentum-space topology. In one of the two generic
universality classes of fermionic quantum vacua the gauge fields, chiral
fermions, Lorentz invariance, gravity, relativistic spin, and other
features of the Standard Model gradually emerge at low energy. The
condensed-matter experience provides us with some criteria for selecting
the proper theories in particle physics and gravity, and even suggests
specific solutions to different fundamental problems. In
particular, it provides us with a plausible mechanism for the solution
of the cosmological constant problem, which I will discuss in some detail.
 }
\end{abstract}

\section{Introduction}

In condensed matter physics we deal with many different strongly correlated and/or strongly interacting
systems. There are no small parameters in such a system and we cannot treat it perturbatively. 
However, from our experience we know that  at  length scales much larger than the inter-atomic spacing,  rather simple behavior emerges which is described by an effective theory. This theory is determined by the universality class  to which the system belongs and does not depend on microscopic details of the system.  There are several types of effective theories.

\begin{itemize}
\item
  A typical example  of an effective theory  is provided by the
Ginzburg-Landau theory describing superconductivity in the vicinity of
the transition temperature $T_c$. This theory, extended to multicomponent
superfluids, superconductors  and Bose condensates, as well as to the
critical phenomena close to $T_c$, is determined by the symmetry of the
system above $T_c$  and describes the symmetry breaking below $T_c$. 
\item
Effective theories of hydrodynamic  type deal with the low-frequency
collective modes away from the critical region. These are the  two-fluid
hydrodynamics of superfluid $^4$He;  the London theory of superconductivity;
their extension to spin and orbital dynamics of superfluid $^3$He; 
elasticity theory in crystals, etc.   This type of effective theories
also describes  topologically non-trivial configurations (including the
topological defects -- singularities of the collective fields protected
by topology, such as quantized vortices) and their dynamics (see the book
\cite{ThoulessBook} for review on the role of the topological quantum
numbers in physics).

  \item
  In the limit $T\rightarrow 0$  an effective quantum field theory
(QFT) emerges. It deals with the ground state of the system (the quantum
vacuum), quasiparticle excitations above the vacuum (analog of elementary
particles),  and their interaction with low-energy collective modes
(bosonic fields).  The QFT kind of effective theories includes the Landau
Fermi-liquid theory with its extension to non-Landau fermionic
systems;  the  quantum Hall effect;   the theory of superfluids and
superconductors at $T\ll T_c$, etc. Here one encounters a phenomenon
which is opposite to the symmetry breaking: the symmetry is enhanced in
the limit $T\rightarrow 0$ \cite{Chadha}. An example is provided by
high-temperature superconductors with gap nodes: close to the nodes
quasiparticles behave as 2+1 Dirac fermions, i.e. their spectrum acquires
the Lorentz invariance.  In superfluid $^3$He-A other elements of the 
relativistic QFT (RQFT) emerge  at $T\rightarrow 0$:  chiral (Weyl) fermions,
gauge invariance, and even some features of effective gravity
\cite{Book}.  

\end{itemize}

In most cases effective  theories cannot be derived from  first
principles, i.e. from the underlying microscopic theory
\cite{LaughlinPines}. If we want to check that our principles of
construction of effective theories are correct and also to search for
other possible universality classes, we  use some very simple models,
which either contain a small parameter, or are exactly solvable. Example is
the BCS theory of a weakly interacting Fermi gas, from which all the types
of the effective theories of superconductivity -- Ginzburg-Landau, London
and QFT -- can be derived within their regions of applicability. 

In particle physics  effective  theories are also major tools
\cite{Ecker}. The Standard Model of quark and leptons and electroweak and
strong interactions operating below 10$^3$GeV is considered as an effective
low-energy RQFT emerging well below the "microscopic" Planck energy scale
$E_{\rm P}\sim 10^{19}$GeV.  It is supplemented by the Ginzburg-Landau type 
theory of electroweak phase transition, and by the hydrodynamic type theory of gravity --
the Einstein general relativity theory. The chiral symmetry and nuclear physics
are the other examples of effective theories;  they emerge in the
low-energy limit of the quantum chromodynamics. In addition,  the
condensed matter examples ($^3$He-A in particular) suggest that not only
these effective theories, 
but even the fundamental physical laws on which they are based (relativistic
invariance, gauge invariance, general relativity, relation between spin and
statistics, etc.)  can be emergent. According to this view the quantum vacuum
-- the modern ether -- can be thought of as some kind of condensed-matter
medium.  This may or may not be true, but in any case it is always
instructive to treat the elementary particle physics with the  methods and
experiences of  the condensed matter physics.

\section{Fermi point and  Standard Model}

\begin{figure}
  \includegraphics[height=0.4\textheight]{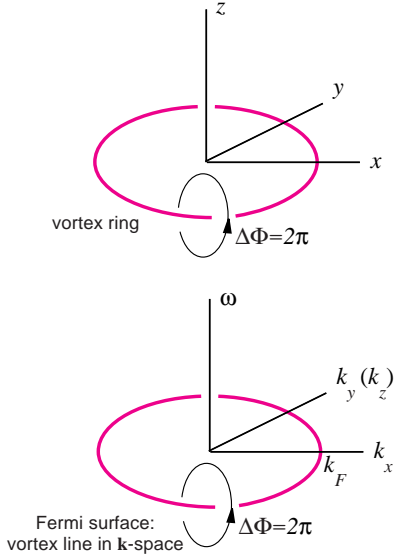}
  \caption{{\it Top}: vortex loop in superfluids and superconductors. The
phase $\Phi$ of the order parameter $\Psi=|\Psi|e^{i\Phi}$ changes by
$2\pi$ around the vortex line and is not determined at the line. {\it
Bottom}: a Fermi surface is a vortex in momentum space. The Green's
function near the Fermi surface is
$G=(i\omega -v_F(k-k_F))^{-1}$. Let us consider the two-dimensional
(2D) system, where
$k^2=k_x^2+k_y^2$. The phase
$\Phi$ of the Green's function
$G=|G|e^{i\Phi}$ changes by
$2\pi$ around the line situated at $\omega=0$ and $k=k_F$ in
the 3D momentum-frequency space $(\omega,k_x,k_y)$. In the 
3D system, where
$k^2=k_x^2+k_y^2+k_z^2$, the vortex line becomes the surface in the
4D momentum-frequency space $(\omega,k_x,k_y,k_z)$ with the
same winding number.
 }
\label{FermiSurface}
\end{figure}

The universality classes of QFT are based  on the topology in momentum
space.  All the information is encoded in the  low-energy asymptote of
the  Green's function for fermions $G({\bf k},i\omega)$. The
singularities in the Green's function in momentum space remind the
topological defects living in real space  \cite{Book,Horava}. Such
a singularity in the ${\bf k}$-space as the Fermi surface is analogous to a
quantized vortex in the ${\bf r}$-space. It is described by the same
topological invariant -- the winding number  (Fig. \ref{FermiSurface}). 
Protected by topology, the Fermi surface survives in
spite of the  interaction between fermions. 
On the emergence of a Fermi surface in string theory see Ref.
\cite{HoravaKeeler}.

Another generic behavior emerges in superfluid $^3$He-A.  The energy
spectrum of the  Bogoliubov--Nambu fermionic quasiparticles in $^3$He-A is
\begin{equation}
E^2 ({\bf k}) = v_F^2(k-k_F)^2+\Delta^2({\bf k})~~,~~
      \Delta^2({\bf k})=     c_\perp^2\,\left({\bf k}\times \hat{\bf
l}\,\right)^2 ,
\label{BogoliubovNambuE}
\end{equation}
where $p_F$ is the Fermi momentum, $v_F$ is the Fermi velocity, and $\hat{\bf l}$
is the direction of the angular momentum  of the Cooper pairs.

\begin{figure}
  \centerline{\includegraphics[width=\linewidth]{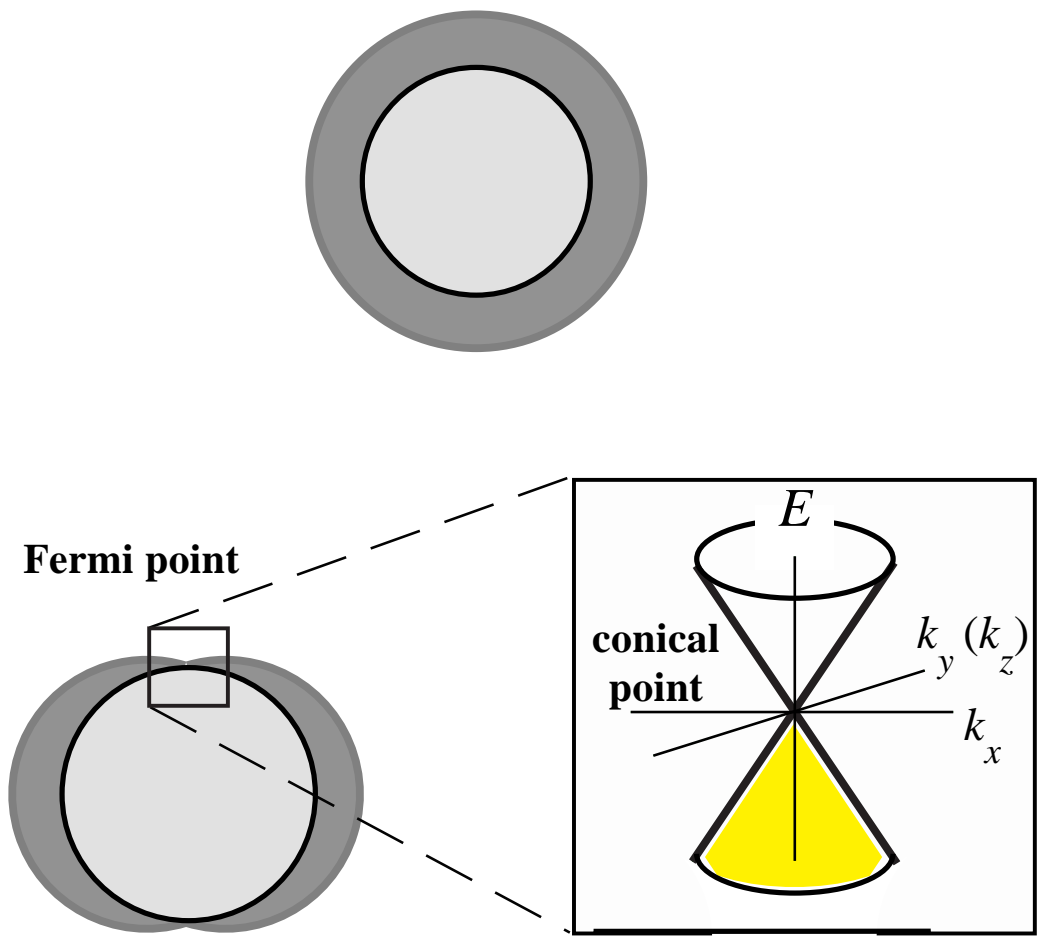}}
  \caption{{\it Top}: isotropic gap in an $s$-wave superconductor.
{\it Bottom left}: in $p$-wave superfluid $^3$He-A  the gap  is
anisotropic and vanishes for
${\bf k}\parallel\hat{\bf l}$. The energy
spectrum (\ref{BogoliubovNambuE}) has two point nodes -- Fermi points.
{\it Bottom right}: close to the node the spectrum
(\ref{BogoliubovNambuRelat}) is similar to the conical spectrum of
right-handed or left-handed fermions of the Standard Model.
 }
\label{GapNode}
\end{figure} 

\begin{figure}
  \includegraphics[height=0.3\textheight]{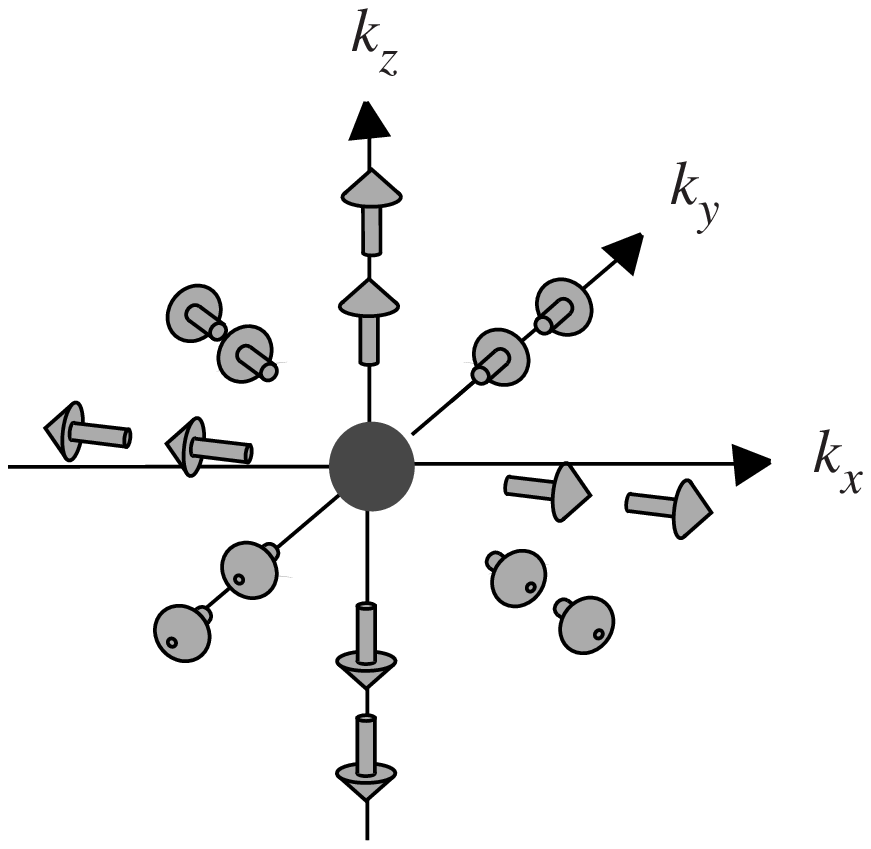}
  \caption{Fermi point is the hedgehog in momentum space. The Hamiltonian
of the fermionic quadiparticles living close to the Fermi point is the
same as either the Hamiltonian for right-handed particles $H=\hbar
c~{\mbox{\boldmath$\sigma$}}\cdot {\bf k}$ or that for the left-handed
particles $H=-\hbar
c~{\mbox{\boldmath$\sigma$}}\cdot {\bf k}$. For each momentum
${\bf k}$ we draw the direction of the particle spin {\boldmath$\sigma$},
which for right-handed particles is oriented along the momentum
${\bf k}$.  The spin distribution in momentum space looks
like a hedgehog, whose spines are represented by spins. The spines point
outward for the right-handed particls, while for the left-handed particles
for which spin is anti-parallel to momentum the spines of the hedgehog
point  inward. Direction of spin is not determined at singular point  ${\bf
k}=0$ in the momentum space. The topological stability of the hedgehog 
singularity under deformations provides the generic behavior of the system
with Fermi points in the limit of low energy. This is the reason why the
chiral particles are protected in the Standard Model and why superfluid
$^3$He-A shares many properties of the vacuum of the Standard Model.
 }
\label{FermiPoint}
\end{figure}

As distinct
from conventional superconductors with
$s$-wave pairing, the gap $\Delta$ in this $p$-wave superfluid is
anisotropic and vanishes for ${\bf k}\parallel\hat{\bf l}$ (Fig.
\ref{GapNode}). As a result the energy spectrum $E({\bf k})$ has zeroes 
at two points ${\bf k}=\pm k_F
\hat{\bf l}$. Such point nodes in the quasiparticle spectrum are
equivalent to  point defects in real space -- the hedgehogs -- and thus
are protected by topology. Moreover the spectrum of elementary particles
in the Standard Model has also the same kind of topologically protected
zeroes  (Fig. \ref{FermiPoint}). The quarks and leptons above the
electroweak transition are massless, and their spectrum $E^2 ({\bf k})
=c^2k^2$ has a zero at
${\bf k}=0$ described by the same topological invariant as the point nodes
in  $^3$He-A. This is the reason why  superfluid $^3$He-A shares many
properties of the vacuum of the Standard Model.

Close to the zeroes the spectrum (\ref{BogoliubovNambuE}) acquires  the
``relativistic''  form:
\begin{equation}
E^2 ({\bf k}) = c_\parallel^2(k_z\pm k_F)^2+  c_\perp^2k_x^2 +
c_\perp^2k_y^2~~,~~c_\parallel\equiv v_F~,
\label{BogoliubovNambuRelat}
\end{equation}
where the $z$-axis is chosen along $\hat{\bf l}$.  For an experimentalist
working with $^3$He-A at low temperature, quasiparticles in
Eq.~(\ref{BogoliubovNambuE}) look like one-dimensional: they move only
along the direction of the nodes (along $\hat{\bf l}$); otherwise they are
Andreev reflected \cite{GreavesLeggett}. A more accurate consideration
in the vicinity of the node in Eq.~(\ref{BogoliubovNambuRelat}) reveals
that they can move in the transverse direction too but about thousand
times slower: the velocity of propagation in the transverse
direction $c_\perp\sim 10^{-3}c_\parallel$. 

On the other hand,  low-energy inner observers  living in  the
$^3$He-A vacuum  would not notice this huge anisotropy. They would find
that their massless elementary particles move in all directions with the
same speed, which is also the speed of light. The reason for this is that
for their measurements of distance they would  use rods made of
quasiparticles: this is their matter.  Such rods are not rigid and their
lengths depend on the orientation. Also, the inner observers would not notice the ``ether
drift'', i.e. the motion of the superfluid vacuum: 
Michelson--Morley-type  measurements of the speed of ``light''  in moving
``ether'' would give a negative result. This resembles the physical
Lorentz--Fitzgerald contraction of length  rods and the physical Lorentz
slowing down of  clocks. Thus the inner observers would finally
rediscover the fundamental Einstein principle of special relativity in
their Universe, while we know that this Lorentz invariance is the
phenomenon emerging at low energy only.

 The physics emerging in the vicinity of the point  nodes is remarkable.
In addition to the Lorentz invariance, the other phenomena of the
RQFT are reproduced. The collective motion of  $^3$He-A
cannot  destroy the topologically protected nodes, it can only shift
the position of the nodes and the slopes of the ``light cone''. The resulting
general deformation of the energy spectrum near the nodes can be written
in the form
 \begin{equation}
g^{\mu\nu}(k_{\mu} - eA_{\mu}^{(a)})(k_{\nu} -
eA_{\nu}^{(a)})=0~.
\label{GeneralDeformation}
\end{equation}
Here the four-vector $A_{\mu}$ describes the degrees of freedom  of the
$^3$He-A vacuum which lead to the shift of the nodes. This is the dynamical
``electromagnetic''  field emerging at low energy, and $e=\pm 1$ is the
``electric''  charge of particles living in the vicinity of north and
south poles correspondingly. The elements of the matrix  $g^{\mu\nu}$
come from the other collective degrees of freedom which form the
effective metric and thus play the role of emerging dynamical gravity.
These emergent phenomena are background independent, if the system stays
within the Fermi-point universality class. Background independence is the
main criterion for the correct quantum theory of gravity.
\cite{Smolin}

One may try to construct a condensed matter system with a large number
of point nodes in the spectrum which would reproduce all the elements of
the Standard Model: 16 chiral fermions per generation; $U(1)$, $SU(2)$
and $SU(3)$ gauge fields; and gravity. There are many open problems on
this way especially with gravity:  in $^3$He-A the equations for the
``gravitational field'' (i.e. for the metric $g^{\mu\nu}$) only
remotely resemble Einstein's equations; while the equation for the
``electromagnetic''  field
$A_{\mu}$ coincides with Maxwell's equation only in a logarithmic
approximation. However, even in the absence of exact correspondence
between the condensed matter system and the Standard Model, there are many common points
which allow us to make  conclusions concerning some unsolved problems in
particle physics and gravity. One of them is the problem of the weight of
the vacuum -- the cosmological constant problem
\cite{Weinberg,Padmanabhan}. 



\section{Vacuum energy and cosmological constant}
\subsection{Cosmological Term and Zero Point Energy}

In 1917, Einstein proposed the model of our Universe with geometry of  a
three-dimensional sphere \cite{einstein}.  To obtain this perfect  Universe, static and homogeneous, as a solution of equations of general relativity, he
added the famous cosmological constant term -- the $\lambda$-term.  At that
time the $\lambda$-term was somewhat strange, since it described the gravity
of the empty space: the empty space
gravitates as a medium with  energy density $\epsilon= \lambda$ and pressure
$p= -\lambda$, where $\lambda$ is the cosmological constant. This medium has
an equation of state
\begin{equation}
p=-\epsilon= -\lambda~.
\label{EOS}
\end{equation} 
When it became clear that our Universe was not static, Einstein removed 
the $\lambda$-term from his equations.

However, later with development of quantum fields it was recognized that  
even in the absence of real particles the space is not empty: the vacuum
is filled with zero point motion which has energy, and according to
general relativity, the energy must gravitate.  For example, each mode of
electromagnetic field with momentum ${\bf k}$ contributes to the vacuum
energy the amount $\frac{1}{2} \hbar \omega({\bf k})=\frac{1}{2} \hbar c
k$. Summing up all the photon modes and taking into account two
polarizations of photons one obtains the following contribution to the
energy density of the empty space and thus to $\lambda$:
\begin{equation}
\lambda=\epsilon_{\rm zero~point}=\int \frac{d^3k}{(2\pi)^3} \hbar ck~.
\label{VacuumEnergyPlanck1}
\end{equation} 

Now it is non-zero, but it is too big,  because it diverges at large $k$.
The natural cut-off is provided by the Planck length scale $a_{\rm P}$,
since the effective theory of gravity -- the Einstein general relativity
-- is only applicable at $k > 1/a_{\rm P}$.  Then  the estimate of the
cosmological constant, $\lambda \sim \hbar c/ a^4_{\rm P}$ exceeds by 120
orders of magnitude the upper limit posed by astronomical observations.

\begin{figure}
  \includegraphics[height=0.25\textheight]{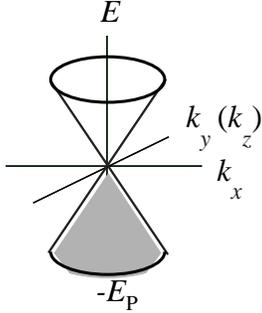}
  \caption{Occupied negative energy levels in the Dirac vacuum produce
a huge negative contribution to the vacuum energy and thus to the
cosmological constant. Summation of all negative energies $E({\bf
k})=-\hbar ck$ in the interval 
$0<E<-E_{\rm P}$, where $E_{\rm P}$ is the Planck energy scale, gives
the energy density of the Dirac vacuum: $\epsilon_{\rm Dirac~vacuum}=-\int
(d^3k/(2\pi)^3)
\hbar ck
\sim  -   \hbar c/ a^4_{\rm P}$, where $a_{\rm P}=\hbar c /E_{\rm P}$ is
the Planck length.
 }
\label{DiracVacuum}
\end{figure} 

There are also contributions to the vacuum  energy from the zero point
motion of other bosonic fields, and a contribution from the
occupied negative energy states of fermions (Fig.  \ref{DiracVacuum}).  If
there is a supersymmetry  -- the symmetry between fermions and bosons --
the contribution of bosons would be  canceled by the negative
contribution of fermions. However,  since the supersymmetry is not exact
in our Universe, it can reduce the discrepancy between theory and
experiment only by about 60 orders of magnitude. The physical vacuum
remains too heavy, and this poses the main cosmological constant problem.  

One may argue that there must exist some unknown but very simple
principle, which leads to nullification of the cosmological constant.
Indeed, in theories in which gravity emerges from the quantum matter
fields, the flat space with $\lambda=0$ appears as a classical equilibrium
solution of the underlying microscopic equations \cite{Amati}.  But what
to do with our estimation of the zero point energy of quantum fields and the
energy of the Dirac vacuum, which
are huge irrespective of whether the vacuum is in equilibrium or not? 

Recently the experimental evidence for non-zero $\lambda$ was 
established: it is on the order of magnitude of the energy  density of
matter,
$\lambda\sim 2-3\epsilon_{\rm matter}$
\cite{Spergel}. People find it easier to believe that the
unknown mechanism of cancellation, if existed, would
 reduce $\lambda$ to exactly zero rather than the observed very low value.  
So, why is $\lambda$   non-zero?  And also, why is it on the
order of magnitude of the matter density? None of these questions
has an answer within the effective quantum field theory, and that is why
our condensed matter experience is instructive, since we know both the
effective theory and the underlying microscopic physics, and are able to 
connect them.

Since we are looking for the general principles governing  the energy
of the vacuum, it should not be of importance for us whether the QFT
is fundamental or emergent. Moreover, we expect that these principles
should not depend on whether or not  the QFT obeys all the
symmetries of the RQFT: these symmetries (Lorentz and
gauge invariance, supersymmetry, etc.) still did not help us to
nullify the vacuum energy.  That is why, to find these principles, we
can look at the quantum vacua whose microscopic structure is well
known at least in principle. These are the ground states of the
quantum condensed-matter systems such as superfluid liquids,
Bose-Einstein condensates in ultra-cold gases, superconductors, insulators, systems
experiencing the quantum Hall effect, etc. These systems provide us
with a broad class of Quantum Field Theories  which are not
restricted by Lorentz invariance. This allows us to consider the
cosmological constant problems from a
more general perspective.

\subsection{Zero Point Energy in Condensed Matter}

The principle which leads to the cancellation of zero-point energy is more
general; it comes from a thermodynamic  analysis
which is not constrained by symmetry or a universality class. 
To see it, let us consider two quantum vacua: the ground states of two
quantum liquids, superfluid $^4$He and one of the two superfluid
phases of  $^3$He, the A-phase. We have chosen these two liquids
because the spectrum of quasiparticles playing the major role at
low energy is ``relativistic''. This allows us to make the connection to
the RQFT. In superfluid $^4$He the relevant  quasiparticles 
are phonons (the quanta of sound waves), and their spectrum is $E(k)=\hbar
ck$, where $c$ is the speed of sound. In superfluid $^3$He-A  the
relevant  quasiparticles  are fermions.  The corresponding ``speed of
light'' $c$ (the slope in the linear spectrum of these fermions in
Eq.~(\ref{BogoliubovNambuRelat})) is anisotropic; it depends on the
direction of their propagation.

Lets us start with superfluid $^4$He and apply the same reasoning as
we did in the case of the electromagnetic field, i.e. we assume that the
energy  of the ground state of the liquid comes from the zero point motion
of the phonon field. Then according to  Eq. (\ref{VacuumEnergyPlanck1}) one
has for the energy density
\begin{equation}
\epsilon_{\rm zero~point}=\frac{1}{2}\int \frac{d^3k}{(2\pi)^3} \hbar ck
\sim 
\frac{\hbar c}{ a^4_{\rm P}} \sim \frac {E_{\rm P}^4}{\hbar^3c^3}  ~,
\label{4He}
\end{equation} 
where the role of the Planck length $a_{\rm P}$ is played by the
interatomic spacing, and the role of the Planck energy scale $E_{\rm
P}=\hbar c/a_{\rm P}$ is provided by the Debye temperature,
$E_{\rm P}= E_{\rm Debye}\sim 1$ K; $c\sim 10^4$ cm/s.

The same reasoning for the  fermionic liquid  $^3$He-A suggests that the
vacuum energy comes from the  
Dirac sea of ``elementary particles'' with spectrum
(\ref{BogoliubovNambuRelat}), i.e. from the occupied levels with negative
energy (see Fig.  \ref{DiracVacuum}):
 \begin{equation}
\epsilon_{\rm Dirac~vacuum}=- 2\int \frac{d^3k}{(2\pi)^3} E({\bf k})
\sim  -\frac { E_{\rm P}^4}{\hbar^3c_\parallel c_\perp^2}~.
\label{3He}
\end{equation} 
Here the Planck energy cut-off is provided by the gap amplitude,
$E_{\rm P}= \Delta \sim c_\perp p_F\sim 1$ mK; 
$c_\parallel\sim 10^4$ cm/s;
$ c_\perp \sim 10$ cm/s.

The above estimates were obtained by using the effective QFT for the 
``relativistic''
fields in the two liquids in the same manner as we did for the quantum
vacuum of the Standard Model. Now let us consider what the exact microscopic
theory tells us about the vacuum energy.

\subsection{Real Vacuum Energy in Condensed Matter} 

The underlying microscopic physics of these two liquids is the physics of
a system of $N$ atoms obeying the conventional quantum mechanics  
and described by the $N$-body Schr\"odinger wave function  $\Psi({\bf
r}_1,{\bf r}_2,
\ldots  , {\bf r}_i,\ldots  ,{\bf r}_N)$. 
The corresponding many-body Hamiltonian is
\begin{equation}
{\cal H}= -{\hbar^2\over 2m}\sum_{i=1}^N {\partial^2\over \partial{\bf r}_i^2}
+\sum_{i=1}^N\sum_{j=i+1}^N U({\bf r}_i-{\bf r}_j)~,
\label{TheoryOfEverythingOrdinary}
\end{equation}
where $m$ is the bare
mass of the atom, and  $U({\bf r}_i-{\bf r}_j)$ is the
pair  interaction of the bare atoms $i$ and $j$.
In the thermodynamic limit where the
volume of the system $V\rightarrow
\infty$ and $N$ is macroscopically large, there emerges an equivalent
description of the system in terms of quantum fields, in a procedure
 known as second quantization.  The quantum field in the
$^4$He ($^3$He) system is presented by the bosonic (fermionic)
annihilation operator 
$\psi({\bf x})$. 
The Schr\"odinger many-body Hamiltonian
(\ref{TheoryOfEverythingOrdinary}) becomes the Hamiltonian of the
QFT \cite{AGDbook}:
\begin{eqnarray}
\hat H_{\rm QFT}=\hat H-\mu \hat N=\int d{\bf x}\psi^\dagger({\bf
x})\left[-{\nabla^2\over 2m} -\mu
\right]\psi({\bf x}) \nonumber \\+{1\over 2}\int d{\bf x}d{\bf y}U({\bf
x}-{\bf y})\psi^\dagger({\bf x})
\psi^\dagger({\bf y})\psi({\bf y})\psi({\bf x}).
\label{TheoryOfEverything}
\end{eqnarray}
Here $\hat N=\int d^3x~\psi^\dagger({\bf
x})\psi({\bf x})$ is the operator of the particle number (number of
atoms); $\mu$ is the chemical potential 
-- the Lagrange multiplier introduced to take into account the
conservation of the number of atoms. Putting aside the philosophical
question of what is primary -- quantum mechanics or quantum field theory 
-- let us discuss the vacuum energy.

The energy density of the vacuum in the above QFT is given 
by the vacuum expectation value of fhe Hamiltonian $\hat H_{\rm QFT}$ in
(\ref{TheoryOfEverything}):
\begin{equation}
\epsilon  =\frac{1}{V}\left<\hat H_{\rm QFT}\right>_{\rm vac}~.
\label{VacuumEnergy}
\end{equation}
In this thermodynamic limit one can apply the Gibbs-Duhem relation,
$E-\mu N -TS=-pV$, which at $T=0$  states: 
\begin{equation}
\left<\hat H\right>_{\rm vac}-\mu \left<\hat N\right>_{\rm vac} =-pV~,
\label{Gibbs-DuhemRelation}
\end{equation}
where   $p$ is the pressure. Using Eqs.~(\ref{TheoryOfEverything}) and
(\ref{VacuumEnergy}) one obtains the relation between the pressure and energy
density in the vacuum state:
\begin{equation}
p=-\epsilon~.
\label{EOSGeneral}
\end{equation} 
It is a general property, which follows from thermodynamics, that the
vacuum behaves as a medium with the above  equation of state. Thus it is
not surprising that the equation of state (\ref{EOSGeneral}) is applicable
also to the particular case of the vacuum of the RQFT in
Eq.~(\ref{EOS}). This demonstrates that the problem of the vacuum energy
can be considered from a more general perspective not constrained by
the relativistic Hamiltonians. Moreover,  it is not important whether
gravity emerges or not in the system.

\subsection{Nullification of Vacuum Energy} 

 Let us consider a situation in which the quantum liquid is completely
isolated from the environment. For example, we launch the
liquid in space where it forms a droplet. The evaporation at $T=0$  is
absent in the liquid, that is why the ground state of the droplet
exists. In the absence of external environment
the external pressure is zero, and thus the pressure of the liquid in its
vacuum state is
$p=2\sigma/R$, where $\sigma$ is the surface tension and $R$ the radius of
the droplet.  In the thermodynamic limit where $R\rightarrow \infty$, the
pressure vanishes. Then according to the equation of
state (\ref{EOSGeneral}) for the vacuum, one has $\epsilon  =-p=0$. This
nullification of the vacuum energy occurs irrespective of whether the
liquid is made of fermionic or bosonic atoms.

\begin{figure}
  \includegraphics[height=0.15\textheight]{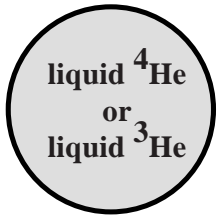}
  \caption{Droplet of quantum liquid. Naive estimation of the vacuum
energy density in superfluid $^4$He as the zero point energy of the phonon
field gives
$\epsilon_{\rm zero~point} 
\sim 
 E_{\rm P}^4/\hbar^3c^3$, where $E_{\rm P}$ is the Debye energy.
 Naive estimation of the vacuum
energy in superfluid $^3$He-A as the energy of the Dirac vacuum gives
$\epsilon_{\rm Dirac~vacuum}
\sim  -  E_{\rm P}^4/\hbar^3c_\parallel c_\perp^2$, where $E_{\rm P}$
is the amplitude of the superfluid gap. But the real energy density of
the vacuum in the droplets is much smaller: for both liquids it is
$\epsilon_{\rm vac}=-2\sigma/R$, where $\sigma$ is the surface tension and $R$
is the radius of the droplet. It vanishes in the thermodynamic limit:
$\epsilon_{\rm vac}(R\rightarrow
\infty)=0$. Inner observers living within the droplet would be surprised
by the disparity of many orders of magnitude between their estimates and
observations. For them it would be a great paradox, which is similar to
our cosmological constant problem.
 }
\label{Droplet}
\end{figure} 

If  observers living within the droplet measure the vacuum energy (or
the vacuum pressure) and compare it with their estimate, Eq. (\ref{4He})
or  Eq. (\ref{3He}) depending on in which liquid they live, they will be
surprised by the disparity of many orders of magnitude between the
estimates and observations (see Fig.  \ref{Droplet}). But we can easily
explain to these observers where their theory goes wrong.  Equations
(\ref{4He}) and  (\ref{3He}) take into account only the degrees of
freedom below the ``Planck'' cut-off energy, which are described by
an effective theory. At higher energies, the microscopic energy of interacting
atoms in Eq.~(\ref{TheoryOfEverything}) must be taken into account, which
the low-energy observers are unable to do. 
When one sums up all the contributions to the vacuum energy, 
sub-Planckian and  trans-Planckian, one obtains the zero result.
The
exact nullification occurs without any special fine-tuning, due to the
thermodynamic relation applied to the whole equilibrium vacuum.

\begin{figure}
  \includegraphics[height=0.3\textheight]{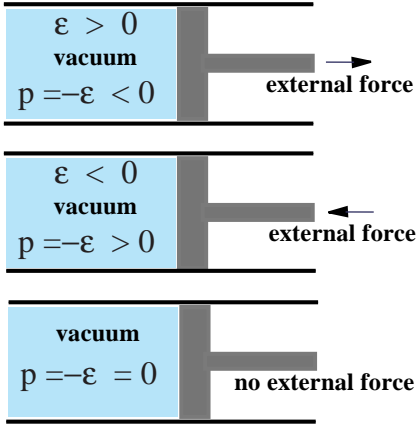}
  \caption{
If the vacuum energy is positive, the vacuum tries to reduce its
volume by moving the piston to the left. To reach an equilibrium, the external
force must be apllied which pulls the piston to the right and compensates for
the negative vacuum pressure. In the same manner, if the vacuum
energy is negative, the applied force  must push the piston to the
left to compensate for the positive vacuum pressure. If there is no
external force from the environment, the self-sustained vacuum must have
zero energy.
 }
\label{Piston}
\end{figure}

This thermodynamic analysis does not depend on the microscopic 
structure of the vacuum and thus can be applied to any quantum vacuum
(Fig. \ref{Piston}), including the vacuum of the RQFT. The
main lesson from condensed matter, which the particle physicists may or may
not accept, is this: the energy density of the homogeneous equilibrium state of the
quantum vacuum is zero in the absence of an external environment. The
higher-energy  (trans-Planckian) degrees of freedom of the quantum vacuum,
whatever they are,  perfectly cancel the huge positive contribution of the
zero-point motion of the quantum fields as well as the huge negative
contribution of the Dirac vacuum. 

This conclusion is supported by the relativistic model, 
in which our world represents the $(3+1)$-dimensional membrane embedded in 
the $(4+1)$-dimensional anti-de Sitter space.  Huge contributions to the  cosmological constant coming from different sources cancel each other without fine-tuning \cite{Andrianov}.
This is the consequence of the vacuum stability.

\subsection{Why the Vacuum Energy is Non-Zero}
\label{non-gravitatingUniverse}

Let us now try to answer the question why, in the present Universe,
the energy density of the quantum vacuum is on the same order of magnitude
as the energy  density of matter. For that let us again
exploit our quantum liquids as a guide. Till now we discussed the pure
vacuum state, i.e. the state without a matter. In the QFT of quantum liquids the
matter is represented by excitations above the vacuum -- quasiparticles.
We can introduce quasiparticles to the liquid droplets by raising their 
temperature $T$ a non-zero value. The quasipartcles in both liquids
are ``relativistic'' and massless. The pressure of the dilute gas of
quasiparticles as a function of $T$ has the same form as the pressure of
ultra-relativistic matter (or radiation) in the hot Universe, if one  
uses the determinant of the effective (acoustic) metric:
\begin{equation}
p_{\rm matter}=\gamma T^4\sqrt{-g}
 ~. 
 \label{RadiationPressure}
\end{equation}
For the quasipartcles in $^4$He, one has
$\sqrt{-g}= c^{-3}$ and  $\gamma=\pi^2/90$; 
for
the fermionic quasiparticles in
$^3$He-A, $\sqrt{-g}= c_\perp^{-2}c_\parallel^{-1}$ and
$\gamma=7\pi^2/360$. 
The gas of quasiparticles obeys the ultra-relativistic 
equation of
state:
\begin{equation}
\epsilon_{\rm matter}=3p_{\rm matter}
 ~.
\label{EOSmatter}
\end{equation}

Let us consider again the droplet of a quantum liquid which
is isolated from the environment, but now at a finite $T$. In the absence
of an environment and for a sufficiently big droplet, where we can neglect
the surface tension, the total pressure in the droplet must be zero. 
This means, that in equilibrium, the  partial pressure of the matter
(quasiparticles) in Eq.~(\ref{RadiationPressure}) must be necessarily
compensated by the negative pressure of the quantum vacuum (superfluid
condensate):
\begin{equation}
p_{\rm matter}+ p_{\rm vac} =0~.
 \label{ZeroPressure}
\end{equation}
The induced negative vacuum pressure leads to the positive vacuum energy
density according the equation of state (\ref{EOSGeneral}) for the
vacuum, and one obtains the following relation between the energy density
of the vacuum and that of the ultra-relativistic matter (or radiation) in thermodynamic equilibrium:
\begin{equation}
\epsilon_{\rm vac} =-p_{\rm vac} =p_{\rm matter}={1\over
3}\epsilon_{\rm matter}~.
\label{VacuumMatterEnergy}
\end{equation} 
This is actually what occurs in quantum liquids, but the resulting
 equation,
\begin{equation}
\epsilon_{\rm vac}
=\kappa \epsilon_{\rm matter}
 ~, 
 \label{kappa}
\end{equation}
with $\kappa=\frac{1}{3}$, does not depend
on the details of the system. It is determined by the equation of state
for the matter and is
equally applicable to:  (i) a superfluid condensate +
quasiparticles with a linear  ``relativistic''  spectrum; and (ii) the vacuum 
 of relativistic quantum fields +
an ultra-relativistic matter (but still in the absence of gravity). 
 
What is the implication of this result to our Universe?  It demonstrates
that when the vacuum is disturbed, the vacuum pressure responds to
the perturbation; as a result the vacuum energy density becomes non-zero.
 In the above quantum-liquid examples the
vacuum is perturbed by a ``relativistic matter''. 
The vacuum is also perturbed by the surface tension of the curved 2D
surface of the droplet which adds its own partial pressure. The
corresponding response of the vacuum pressure
 is $2\sigma/R$.
 
Applying this to the general relativity, we can conclude that
 the homogeneous
equilibrium state of the quantum vacuum without a matter is not
gravitating, but the disturbed quantum vacuum has
a weight. In the Einstein Universe the vacuum
is perturbed by the matter and also by the gravitational
field (the 3D space curvature). These perturbations induce 
the non-zero cosmological constant, which was first calculated by
Einstein who found that $\kappa=\frac{1}{2}$ for the
cold static Universe \cite{einstein} (for the hot static Universe filled
with ultrarelativistic matter, $\kappa=1$). In the expanding or rotating
Universe the vacuum is perturbed by expansion or rotation, etc. In all
these cases, the value of the vacuum energy density is proportional to
the magnitude of perturbations. Since all the perturbations of the vacuum
are small in the present Universe, the present cosmological constant must
be small. 

\section{Conclusion}

 What is the condensed matter experience good for? It provides us with
 some criteria for selecting
the proper theories in particle physics and gravity, 
For example, some scenarios of inflation are prohibited, since 
according to the Gibbs-Duhem relation the metastable false vacuum also
has  zero energy \cite{VolovikAnnals}. The condensed matter experience suggests its
specific solutions to different fundamental problems, such as
cosmological constant problem. It demonstrates how the symmetry and
physical laws emerge in different corners of parameters, including the
zero energy corner. It also provides us with a variety of universality
classes and corresponding effective theories, which are not restricted by
Lorentz invariance and by other imposed symmetries.  

The effective field theory is the major tool in condensed matter 
and particle physics. But it is not appropriate for the calculation of
the vacuum energy in terms of the zero-point energy of effective quantum
fields. Both in condensed matter and particle physics, the contribution of
the zero-point energy to the vacuum energy exceeds, by many orders of magnitude,
 the measured vacuum energy.  The condensed matter, however, gives
a clue to this apparent paradox: it demonstrates that this huge
contribution is cancelled by the microscopic (trans-Planckian) degrees of
freedom that are beyond the effective theory. We may know nothing about the
trans-Planckian physics, but the cancellation does not depend on the
microscopic details, being determined by the general laws of
thermodynamics. This allows us to understand, in particular, what happens
after the cosmological phase transition, when the vacuum energy decreases
and thus becomes negative. The microscopic degrees of freedom will
dynamically readjust themselves to the new vacuum state,  relaxing the
vacuum energy back to zero \cite{VolovikAnnals}. Actually, the observed
compensation of zero-point energy suggests that there exists an underlying
microscopic background and the general relativity is an effective
theory rather than a fundamental one. 

In the disturbed vacuum, the compensation is not complete, and this
gives rise to the non-zero vacuum energy proportional to disturbances. 
The cosmological constant is small simply because  in the present
Universe all the disturbances are small: the matter is very dilute, and
the expansion is very slow, i.e. the vacuum of the Universe is very close
to its equilibrium state. One of the disturbing factors in our Universe is
the gravitating matter, this is why it is natural that the measured
cosmological constant is on the order of the energy density of the matter:
$\kappa\sim 3$ in Eq.~(\ref{kappa}).

Thus, from the condensed matter point of view, there are no great
paradoxes related to the vacuum energy and cosmological constant. Instead
we have the practical problem to be solved: how to calculate $\kappa$ and
its time dependence. Of course, this problem is not simple, since it
requires the physics beyond the Einstein equations, and there are too
many routes on the way back from the effective theory to the microscopic
physics.


  
This work is supported in part by the Russian Ministry of
Education and Science, through the Leading Scientific School
grant $\#$2338.2003.2, and by the European Science Foundation 
COSLAB Program.





\bibliography{sample}

\begin{thebibliography}{99}

\bibitem{ThoulessBook} D.~J. Thouless, \emph{Topological Quantum Numbers in
Nonrelativistic Physics}, World Scientific,  Singapore, 1998.


\bibitem{Chadha} S. Chadha  and  H.~B. Nielsen,  \emph{Nucl. Phys.},  \textbf{B~217}, 125--144 (1983).


\bibitem{Book} G.~E. Volovik, \emph{The Universe in a Helium
Droplet}, Clarendon Press,  Oxford, 2003.

\bibitem{LaughlinPines} R.~B. Laughlin and D. Pines,   \emph{Proc. Natl Acad. Sci. USA}, \textbf{97}, 28--31
(2000).

\bibitem{Ecker} G. Ecker, ``Effective Field Theories'', hep-ph/0507056.

\bibitem{Horava}  P. Horava,  \emph{Phys. Rev. Lett.}, \textbf{95},
 016405 (2005).


\bibitem{HoravaKeeler} P. Horava and C.~A. Keeler,
 ``Noncritical M-Theory in 2+1 Dimensions as a Nonrelativistic
Fermi Liquid'', hep-th/0508024.
      

\bibitem{GreavesLeggett} N.~A. Greaves and A.~J. Leggett,  \emph{J. Phys. C: Solid State Phys.},  \textbf{16}, 4383--4404 (1983).

\bibitem{einstein}
A.~Einstein, 
 \emph{Sitzungberichte der Preussischen Akademie der Wissenschaften},
 \textbf{1}, 142--152 (1917); also in a translated version in  \emph{The
Principle of Relativity},  Dover, 1952.

\bibitem{Smolin}  L. Smolin,  
``The Case for Background Independence,'' hep-th/0507235.

\bibitem{Weinberg}  S. Weinberg,  \emph{Rev. Mod. Phys.},
  \textbf{61}, 1--23 (1989).

\bibitem{Padmanabhan}
 T. Padmanabhan,  
  \emph{Phys. Rept.},  \textbf{380},  235--320 (2003).

\bibitem{Spergel} D.~N. Spergel, L. Verde, H.~V. Peiris,  {\it et al.}, 
 \emph{Astrophys. J. Suppl.},  \textbf{148}, 175--194  (2003).

\bibitem{Amati} D.~Amati and G.~Veneziano, 
 \emph{Phys. Lett.},  \textbf{105~B}, 358--362  (1981).

\bibitem{AGDbook} A.~A. Abrikosov, L.~P. Gorkov, and I.~E. Dzyaloshinskii,
\emph{Quantum Field Theoretical Methods in Statistical Physics},
Pergamon, Oxford,  1965.

\bibitem{Andrianov}
A.~A. Andrianov, V.~A. Andrianov, P. Giacconi, and R. Soldati,
 \emph{JHEP},  \textbf{0507}, 003  (2005).


\bibitem{VolovikAnnals} G.~E. Volovik, \emph{Annalen der Physik}, 
\textbf{14}, 165--176 (2005).





\end{thebibliography}


\end{document}